\documentclass[12pt,preprint]{aastex}      

\usepackage[latin1]{inputenc}
\usepackage{lscape}
\usepackage{amsmath}

\slugcomment{Version 1.01; Printed: \today}

\shorttitle{A Dwarf Protoplanetary Disk around XZ Tau B}

\shortauthors{Osorio et al.}

\begin{document}

 \title{A Dwarf Transitional Protoplanetary Disk around XZ Tau B}


\author{Mayra Osorio\altaffilmark{1}, 
Enrique Mac\'{\i}as\altaffilmark{1},
Guillem Anglada\altaffilmark{1},
Carlos Carrasco-Gonz\'alez\altaffilmark{2},
Roberto Galv\'an-Madrid\altaffilmark{2},
Luis Zapata\altaffilmark{2},
Nuria Calvet\altaffilmark{3},
Jos\'e  F. G\'omez\altaffilmark{1},
Erick Nagel\altaffilmark{4},
Luis F. Rodr\'{\i}guez\altaffilmark{2},
Jos\'e M. Torrelles\altaffilmark{5 \dagger},
Zhaohuan Zhu\altaffilmark{6}
}

 \altaffiltext{1}{Instituto de Astrof\'\i sica de Andaluc\'\i a (CSIC), 
Glorieta de la Astronom\'\i a s/n, E-18008 Granada, Spain;  email: 
{\tt osorio@iaa.es}}
 \altaffiltext{2}{Instituto de Radioastronom\'{\i}a y Astrof\'{\i}sica 
UNAM, Apartado Postal 3-72 (Xangari), 58089 Morelia, Michoac\'an, Mexico}
 \altaffiltext{3}{Department of Astronomy, University of Michigan,  825 
Dennison Building, 500 Church St, Ann Arbor, MI 48109, USA}
 \altaffiltext{4}{Departamento de Astronom\'{\i}a, Universidad de 
Guanajuato, Guanajuato, Gto 36240, Mexico}
 \altaffiltext{5}{Institut de Ci\`encies de l'Espai (CSIC)-Institut de 
Ci\`encies del Cosmos (UB)/IEEC, Mart\'{\i} i Franqu\`es 1, E-08028 
Barcelona, Spain}
 \altaffiltext{6}{Department of Astrophysical Sciences, Princeton 
University, Princeton, NJ 08544, USA}
 \altaffiltext{$\dagger$}{The ICC (UB) is a CSIC-Associated Unit through 
the ICE}

\begin{abstract}

We report the discovery of a dwarf protoplanetary disk around the star XZ 
Tau B that shows all the features of a classical transitional disk but 
on a much smaller scale. The disk has been imaged with the Atacama Large 
Millimeter/Submillimeter Array (ALMA), revealing that its dust emission 
has a quite small radius of $\sim$3.4 au and presents a central cavity 
of $\sim$1.3 au in radius that we attribute to clearing by a compact 
system of orbiting (proto)planets. Given the very small radii involved, 
evolution is expected to be much faster in this disk (observable changes 
in a few months) than in classical disks (observable changes requiring 
decades) and easy to monitor with observations in the near future. From 
our modeling we estimate that the mass of the disk is large enough to 
form a compact planetary system.

\end{abstract}

\keywords{planet-disk interactions --- protoplanetary disks --- stars: 
formation --- stars: individual (XZ Tau B) --- stars: pre-main 
sequence}

\section{Introduction}

Planetary systems originate from the evolution of accretion disks of gas 
and dust that develop around young stars as part of the star formation 
process itself (Blum \& Wurm 2008). However, the details of how planets 
form are far from well understood. Some accretion disks, known as 
``transitional disks'' (Calvet et al. 2005), present central cavities 
and annular gaps in their dust emission that have been attributed to the 
effects of tidal interactions of orbiting planetary or protoplanetary 
bodies (Papaloizou et al. 2007, Zhu et al. 2011, Andrews et al. 2011, 
Osorio et al. 2014) and are considered signposts of the planet formation 
process. Typical transitional disks imaged so far have radii of 50-100 
au and masses of 10-100 $M_J$, with central cavities of 15-70 au in 
radius (Andrews et al. 2011, Espaillat et al. 2014, Andrews 2015). 
Nevertheless, some results, based on the spectral energy distribution 
(SED) modeling, suggest that significantly smaller disks should exist 
(McClure et al. 2008, Pi\'etu et al. 2014), but direct imaging has not 
been possible yet.

XZ Tau B is a young M2 dwarf star (see stellar properties in Table 1) in 
the L1551 molecular cloud. It belongs to a triple system composed of the 
close pair XZ Tau A/C (separation $\sim0.09''$) and XZ Tau B (currently 
located $\sim0.3''$ to the NW; Carrasco-Gonz{\'a}lez et al. 2009). A 
sequence of expanding bubbles imaged by the HST (Krist et al. 2008) has 
been attributed to XZ Tau A/C (Carrasco-Gonz{\'a}lez et al. 2009, Zapata 
et al. 2015), while high velocity jets have been associated with both XZ 
Tau A/C and XZ Tau B (Krist et al. 2008).

During the Long Baseline Campaign of the ALMA Science Verification 
process, a field centered on HL Tau was observed at 2.9 mm and 1.3 mm 
(ALMA Partnership et al. 2015a, 2015b; hereafter AP2015a, AP2015b). XZ 
Tau B was reported only as an unresolved continuum source at 2.9 mm. 
Here we present a detailed analysis of the 1.3 mm continuum observations 
that angularly resolve the source.

\section{Observations}

The observations were carried out between 2014 October 14 and November 
14 using 42 antennas of ALMA, with baselines from 12 to 15,240 m. The 
phase center was at the position of HL Tau, $\alpha$(J2000)=$4^{\rm 
h}31^{\rm m}38.4263^{\rm s}$, $\delta$(J2000)=$18^{\circ}13'57.047''$. 
The 1.3 mm data were obtained from 2014 October 24 to 31 with the ALMA 
correlator configured in 4 spectral windows of 2000 MHz and 128 channels 
each. The 2.9 mm data were obtained using wide spectral windows for the 
continuum and narrow spectral windows centered on the HCO$^+$(1-0), 
HCN(1-0), CO(1-0), and CN(1-0) lines. A description of the observational 
setup and the calibration process is given in AP2015a, AP2015b.

The continuum emission at 2.9 mm was imaged by AP2015a (beam 
$=0.085''\times0.061''$, PA=$-179^{\circ}$; rms = 24 $\mu$Jy 
beam$^{-1}$). We obtained cleaned, continuum subtracted, channel maps 
(channel width $=0.25$ km s$^{-1}$; beam $=0.10''\times0.06''$, 
PA=$12^\circ$) for the observed line transitions. No line emission was 
detected towards XZ Tau.

Images at 1.3 mm were obtained with the task {\em clean} of CASA 
(version 4.2.2). XZ Tau falls $\sim24''$ away from the phase center, 
where the response of the primary beam (FWHM $\simeq27''$ at 1.3 mm) is 
only 1/19. However, the extraordinary sensitivity of ALMA allows a good 
signal-to-noise imaging. To avoid HL Tau sidelobes in the XZ Tau field, 
we first cleaned the HL Tau emission and subtracted it from the uv data.
  We tried several self-calibration strategies. Although 
self-calibration improves slightly the images of HL Tau, it blurs the XZ 
Tau images. We attribute these unfavorable effects on XZ Tau to the lack 
of a strong compact source in the field and to the large separation of 
XZ Tau from the phase center. Since we are interested in XZ Tau, we did 
not apply self-calibration in our final images. Given the narrow channel 
width and small integration time per visibility, the expected bandwidth 
and time smearing are negligible at the position of XZ Tau ($0.0016''$ 
and $0.0035''$, respectively).

Figure 1a shows our primary-beam corrected 1.3 mm image of XZ Tau B. The 
source is angularly resolved, with a size of $\sim0.05''$ and a flux 
density of $7\pm2$ mJy\footnote{Measured in a natural-weight image}. At 
2.9 mm it was reported as angularly unresolved (size $<0.054''$) with a 
flux density of $1.83\pm0.12$ mJy (AP2015a; Zapata et al. 2015). 
Uncertainties in flux densities have been calculated as in AP2015a, but 
adding quadratically the absolute flux density calibration uncertainty 
(5\%) and the primary beam response uncertainty due to pointing errors 
($\sim0.6''$), using the Dzib et al. (2014) prescription.

 Since the 2.9 mm observations are less affected by the primary beam 
attenuation, they are much more sensitive. The fact that the source size 
upper limit set by these observations is similar to the observed size at 
1.3 mm ($\sim0.05''\simeq7$ au) indicates that the sensitivity of the 
1.3 mm image is high enough to reveal the full structure of the source 
and not just the brightest part. Thus, we interpret the observed 
emission as tracing the dust of a very small ($\sim3.5$ au in radius) 
circumstellar disk.

Interestingly, the 1.3 mm ALMA image reveals substructure in the disk 
(Fig.~1a). Emission decreases towards the center, indicating a hole or 
cavity. Otherwise, the emission would peak towards the central position. 
We have plotted the real component of the visibility profile (Fig.~1b), 
which shows the characteristic null and negative region that confirm the 
presence of a central hole in the disk (e.g., Andrews et al. 2009). The 
null falls around 5-8 M$\lambda$, corresponding to hole radii of 0.6-0.9 
au to 1.4-2.2 au for the extreme cases of an infinite disk and a thin 
ring, respectively (Hughes et al. 2007)\footnote{Visibilities have been 
recentered on XZ Tau B, but not deprojected to account for the disk 
inclination. So, these values are only rough estimates}. Since our small 
disk should be something intermediate, we estimate a radius of the hole 
$\sim$1 au, consistent with the value obtained from our modeling of the 
SED and image ($\S3$). Thus, XZ Tau B appears to be a ``transitional 
disk'' (Calvet et al. 2005) with a small central cavity probably due to 
the tidal forces created by an orbiting substellar object or protoplanet 
(Andrews et al. 2011). In order to substantiate this interpretation we 
carried out a detailed modeling.

\section{Modeling}

The disk parameters are determined by modeling and fitting the observed 
SED and the normalized radial intensity profile of the 1.3 mm image. To 
construct the observed SED we compiled photometric and spectroscopic 
data from the Spitzer, WISE, Akari, and IRAS databases, from the 
literature (Hartigan \& Kenyon 2003, White \& Ghez 2001, 
Carrasco-Gonz{\'a}lez et al. 2009, AP2015a, Forgan et al. 2014), and 
from this paper. Measurements that do not separate the A and B 
components have been taken as upper limits.

Our model includes the contributions from both the central star and the 
disk. Since the XZ Tau B star is known to be optically variable (Sandell 
\& Aspin 1998), we reanalyzed the results of Hartigan \& Kenyon (2003) 
but using the photometry of XZ Tau A from Coffey et al. (2004) to 
estimate the aperture correction. The stellar and accretion (veiling) 
luminosities of XZ Tau B were obtained following Pecaut \& Mamajek 
(2013), Kenyon \& Hartmann (1995), and Calvet \& Gullbring (1998) 
assuming an M2 star and an 8000 K blackbody as the veiling source. 
Finally, using the Siess et al. (2000) tracks, the stellar parameters 
were derived (Table 1). The contribution of the central star to the SED 
(Fig.~3a) is calculated by using the reanalyzed fluxes and extrapolating 
to other wavelengths following Kenyon \& Hartmann (1995) and Pecaut \& 
Mamajek (2013).

The disk is modeled using an updated version of the irradiated 
$\alpha$-accretion disk models with dust settling developed by D'Alessio 
et al. (2006). A dust grain population similar to the interstellar 
medium is used in the upper layers of the disk, while in the mid-plane a 
population of larger dust grains, with radii up to 1 mm, is assumed. The 
grain mixture composition is the same as in Osorio et al. (2014), but 
incorporating water ice with the abundance given by McClure et al. 
(2015), resulting in a dust-to-gas ratio of 0.0085. A central cavity is 
included in the model by emptying the innermost regions of the disk. The 
edge of this region, or wall, is directly irradiated by the star and the 
accretion shock and, thus, heated to a higher temperature (D'Alessio et 
al 2005).

The high H, K, and L band fluxes (Hioki et al. 2009, White \& Ghez 2001) 
indicate that hot dust, that may correspond to a residual inner disk, is 
present inside the cavity, suggesting that XZ Tau B is at an earlier 
pre-transitional stage (Espaillat et al. 2008). However, because of the 
variability of the star and the limited data in this wavelength range, 
we cannot determine the properties of this inner component, and thus we 
did not include it in our model.

The disk inclination (angle between the rotation axis and the 
line-of-sight) and PA were fitted by exploring different values, 
assuming as an initial guess that the disk lies in the B and A/C orbital 
plane ($i=47^\circ$; Carrasco-Gonz\'alez et al. in preparation) and is 
perpendicular to the direction of the observed collimated jet (Krist et 
al. 2008).
 
Hence, the main free parameters are the viscosity parameter, $\alpha$, 
the mass accretion rate in the disk, $\dot M_{\rm disk}$, and the degree 
of settling, $\epsilon$. Planet-forming disks are expected to have dust 
populations highly settled onto the midplane (i.e., a low value of 
$\epsilon$, defined as the dust-to-gas ratio in the atmosphere relative 
to the total of the disk), so that planetesimals can grow through the 
aggregation of large grains. Thus, we explored low values of $\epsilon$, 
$0.001\le\epsilon\le0.1$.

To analyze how the viscosity affects the gas and dust evolution, we have 
carried out gas-dust two fluid hydrodynamical simulations as in Zhu et 
al. (2012). With a large viscosity (e.g., $\alpha=10^{-2}$), such a 
small disk evolves very fast. As shown in the right panel of Figure 2, 
the disk gas surface density decreases by 4 orders of magnitude within 
0.1 Myr and all the dust drifts to the central star.  With a smaller 
viscosity (e.g., $\alpha=10^{-3}$ as shown in the left panel of Fig.~2), 
$\sim$0.1 $M_J$ mass planets are sufficient to produce a cavity which is 
almost two orders of magnitude deep.  Thus, multiple planets could 
account for the observed cavity in XZ Tau B as long as the viscosity is 
small enough ($\alpha\le10^{-3}$).

Simulations also show that accretion onto the planet creating the gap 
can account for up to 90\% of $\dot M_{\rm disk}$ (Zhu et al. 2011) and 
would reduce the mass accretion rate onto the star, $\dot M_*$. 
Therefore, we explored values of $\dot M_{\rm disk}$ in the range $\dot 
M_*<\dot M_{\rm disk}<10\,\dot M_*$, where $\dot M_*$ is given in Table 
1.

We have run a grid of 40 models with parameters in the above-mentioned 
ranges. Since we are interested in studying the capability of the disk 
to form planets, we have selected the model that fits the data with the 
highest allowed value for the viscosity parameter, which gives the 
lowest disk mass (9 $M_J$). As we show in $\S4$, even this low-mass disk 
is capable to form a planetary system. The parameters of this disk are 
given in Table 1.

The resulting SED, showing the separate contributions of the main 
components, is plotted in Figure 3a. The free-free contribution from the 
ionized jet has also been taken into account in the fit, showing that it 
is negligible in the mm range. A comparison of the observed and model 
intensity profiles along the major axis of the disk is shown in Figure 
3b. Figure 3c shows the surface density and temperature model profiles. 
Figure 3d shows a CASA simulated image of the model emission at 1.3 mm 
as it would be observed with the same ALMA configuration as Figure 1. 
These figures show that the model reproduces the observations reasonably 
well. Thus, our results support the interpretation that XZ Tau B is 
associated with a dwarf transitional disk.

\section{Discussion}

Modeling shows that the outer radius of the disk is 3.4 au and the 
radius of the central cavity is 1.3 au (Table 1). These radii are well 
constrained by the intensity profile and are much smaller than those of 
other transitional disks (typically $\sim$50-100 au for the disk and 
$\sim$15-70 au for the cavity; Andrews et al. 2011, Espaillat et al. 
2014, Andrews 2015). Tidal interactions in a close binary are expected 
to truncate circumstellar disks to an outer radius $\sim$1/3 of the 
binary separation (Papaloizou \& Pringle 1977). This has been observed 
in the L1551-IRS5 binary system of disks, each 10 au in radius 
(Rodr{\'{\i}}guez et al. 1998). Interestingly, the radius of the XZ Tau 
B disk is significantly smaller than the value of $\sim$14 au expected 
from tidal truncation, given the separation of $\sim$42 au between XZ 
Tau B and the A/C pair. A highly eccentric orbit could truncate the disk 
at a smaller radius, but the analysis of the relative positions of the 
stars over $>$20 yr favors a nearly circular orbit (Carrasco-Gonz\'alez 
et al. 2009, Carrasco-Gonz\'alez et al. in preparation).

The reason for the small size of the XZ Tau B disk is uncertain. It is 
feasible that the disk was originally small; that simple tidal 
truncation models (Papaloizou \& Pringle 1977) may not apply for the 
particular geometry of this disk and truncation occurs at a smaller 
scale; that the disk is outwardly truncated by a forming planet in an 
outer orbit (Osorio et al. 2014); or that the outer parts of the disk 
have been removed by other mechanisms (e.g., swept out by the sequence 
of expanding bubbles from the A/C stars; Krist et al. 2008). It is 
possible that the gas component of the XZ Tau B disk is more extended 
than the dust, as it occurs in standard disks (e.g., HD 163296; de 
Gregorio-Monsalvo et al. 2013). Unfortunately, the sensitivity of our 
line observations is insufficient to set a tight constraint to the gas 
disk size. Anyhow, the disk of dust in XZ Tau B is much smaller than any 
other angularly resolved disk of dust imaged so far.

The observed image (Fig.~1a) is marginally asymmetric, with the flux 
density in the southeast region $\sim30\%$ higher than in the northwest 
one. Such an asymmetry could not result from opacity effects due to the 
disk inclination since our modeled image (Figs.~3b, 3d) is symmetric. 
Instead, this asymmetry is suggestive of a dust trap, where the largest 
dust grains accumulate (e.g., Birnstiel et al. 2013, van der Marel et 
al. 2013). However, this needs to be confirmed with higher sensitivity 
data.

 These results suggest that XZ Tau B shows the features that 
characterize transitional disks, but on a much smaller scale (e.g., 
compare Fig.~1a with Fig.~1b in Osorio et al. 2014). Since the evolution 
of these features is determined by their orbital motions around the 
central star, a dwarf disk like XZ Tau B is expected to evolve 
$\sim$50-500 times faster than their bigger counterparts. Unfortunately, 
the current 1.3 mm ALMA observations, spanning just one week, are 
insufficient to search for disk evolution, but significant changes can 
occur in observations separated by only a few months. Thus, we 
anticipate that the disk in XZ Tau B, and possibly other similar dwarf 
disks, may serve in the near future as valuable small-scale models for a 
fast and efficient study of the evolution of transitional disks.

The diversity of planetary systems observed in the exoplanet surveys 
suggests that an equivalent diversity should be found in their 
progenitors, the protoplanetary disks. In particular, the Kepler mission 
has identified a number of ``low-mass compact multiple-planet systems'', 
orbiting within $<1$ au from the star and with planetary masses ranging 
from a fraction to a few times the Earth's mass (Lissauer et al. 2011, 
2014, Jontof-Hutter et al. 2015). Dwarf disks similar to that found in 
XZ Tau B appear as the natural precursors of these systems.

 To fit both the relatively high mm flux density of the disk and its 
small size, a high mass accretion rate was needed in our model. This, 
combined with the low viscosity of the disk, resulted in very high disk 
surface densities (Fig.~3c). These values are higher than those expected 
at the inner regions of larger disks around this type of stars (Williams 
\& Cieza 2011, Andrews 2015) but they are still one order of magnitude 
smaller than the values of the ``minimum mass protoplanetary nebula'' 
estimated by Swift et al. (2013) to form ``in situ'' the Kepler-32 
planetary system.

It is also interesting to compare with the compact system of five 
sub-Earth radius planets around the K0 star Kepler-444A, in a 
hierarchical triple stellar system. Dupuy et al. (2016) estimated a 
small radius of $\sim$2 au and a relatively large mass $\gtrsim70\,M_J$ 
for the primordial protoplanetary disk around Kepler-444A. These authors 
propose that the outer regions of such a massive disk would have been 
unstable, leading to the formation of the triple star system through 
gravitational fragmentation. XZ Tau B could be similar to Kepler-444A. 
However, the disk of XZ Tau B presents a central gap, while the planets 
around Kepler-444A have masses well below the gap-opening mass (Dupuy et 
al. 2016). This implies that the planets recently formed in XZ Tau B are 
probably more massive than the planets around Kepler-444A. The reason of 
this could be the difference in the position of the snowline. Whereas 
the whole disk of Kepler-444A would have fallen within its snowline, our 
model of XZ Tau B shows that (for points slightly above the disk 
midplane, where the minimum temperature is reached) it could be located 
inside the cavity, at a radius of $\sim$0.5 au, where $T_{\rm min}=180$ 
K (Fig.~3c). Therefore, giant planets, which form much more easily 
beyond the snowline (e.g., Ros \& Johansen 2013), could be forming in 
the XZ Tau B disk cavity.

Some theoretical studies already pointed to the possible existence of a 
relatively large population of very small disks
 (McClure et al. 2008, Pi{\'e}tu et al. 2014, Kraus et al. 2015, Furlan 
et al. 2016). However, none of these putative dwarf disks has been 
angularly resolved. In XZ Tau B we have been able, not only to angularly 
resolve the disk and determine its size, but also to observe and model 
its substructure at au-scales. The Kepler mission raised a number of 
puzzling questions regarding the observed planetary systems at distances 
$<1$ au from the star, such as the debate of migration versus ``in 
situ'' planetary formation (Ogihara et al. 2015), or the abundance of 
super-Earths very close to the star (Lee et al. 2014). XZ Tau B opens a 
new window to investigate with ALMA observations the disk evolution and 
first stages of planet formation on time-scales and at radii that so far 
remained unexplored.


 \begin{figure}
 \begin{center}
\hspace{-1.5cm}\includegraphics[height=10.5cm]{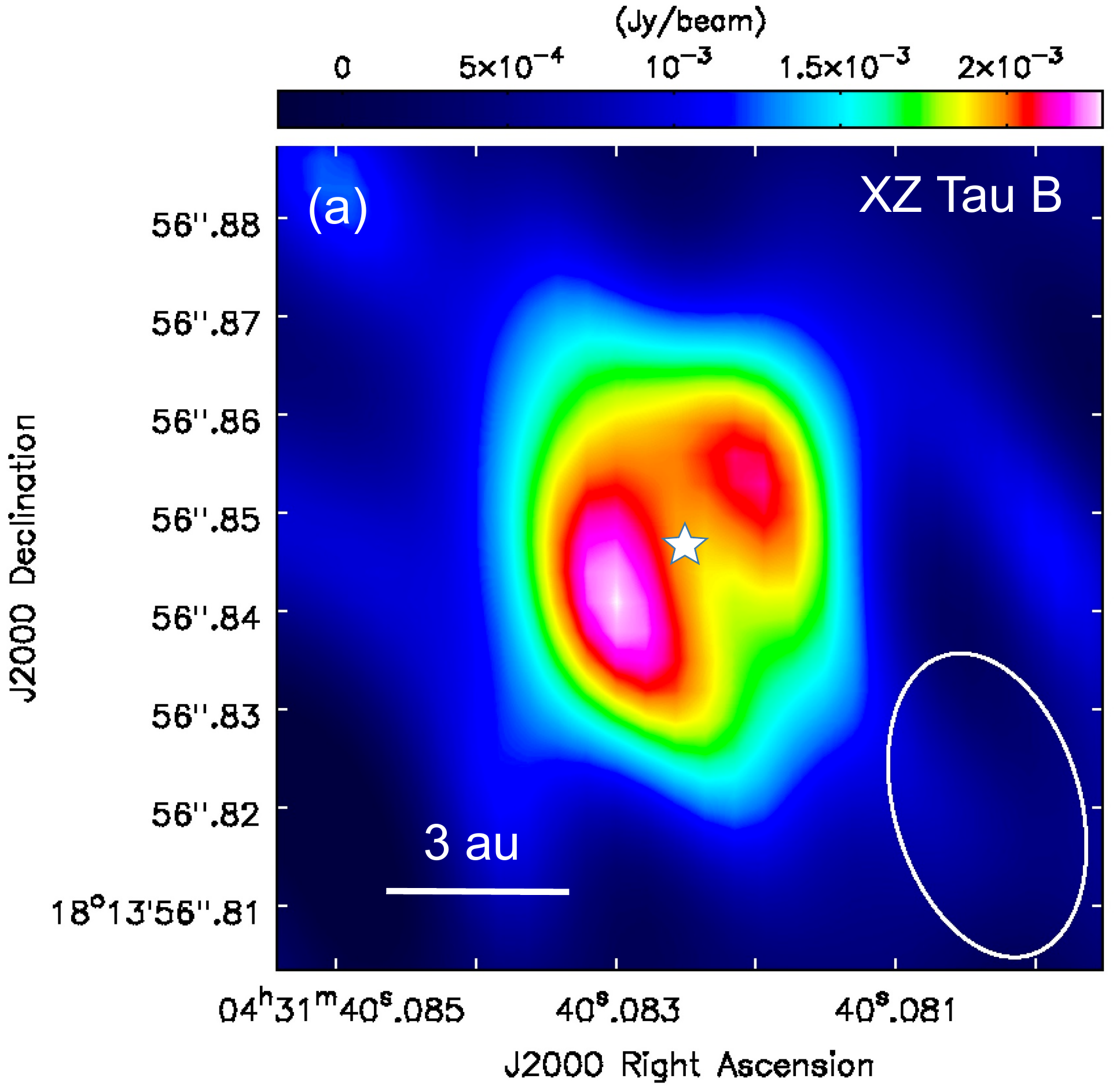}
\includegraphics[height=6.5cm]{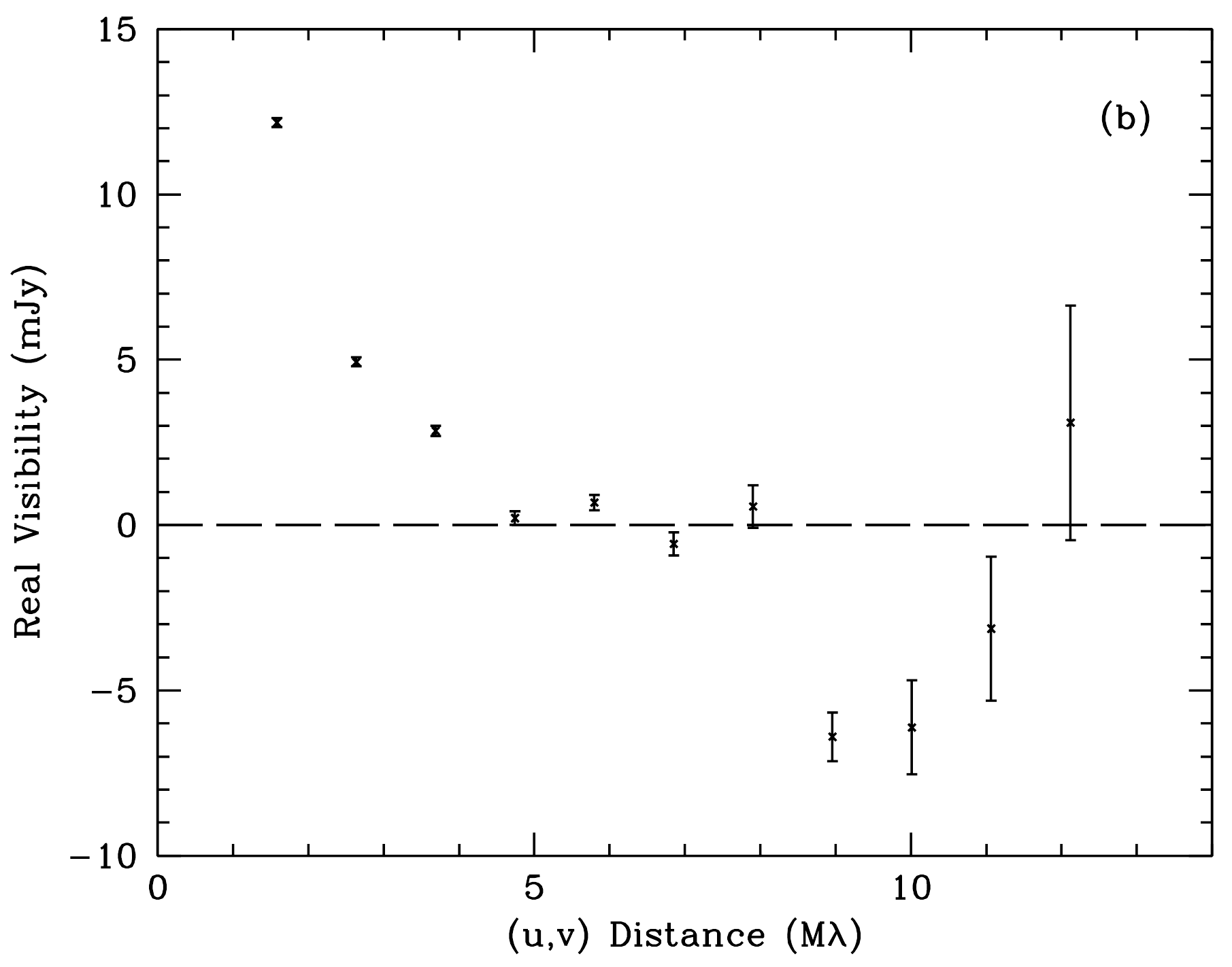}
 \caption{(a) ALMA image at 1.3 mm (robust=0, rms=0.28 mJy beam$^{-1}$) 
of the disk of dust around XZ Tau B (indicated by a star, at 
$\alpha$(J2000)=$4^{\rm h}31^{\rm m}40.0825^{\rm s}$, 
$\delta$(J2000)=$18^{\circ}13'56.847''$). The synthesized beam 
($0.032''\times0.019''$, PA=16$^\circ$) is shown in the lower-right 
corner. The intensity decrease in the northeast and southwest edges is 
an observational effect due to the beam elongation along this direction 
(see Osorio et al. 2014). (b) Azimuthally averaged visibility profile 
plotted in $\sim$1 M$\lambda$ bins.}
 \label{Fig1}
 \end{center}
 \end{figure}

\clearpage

 \begin{figure}
\plotone{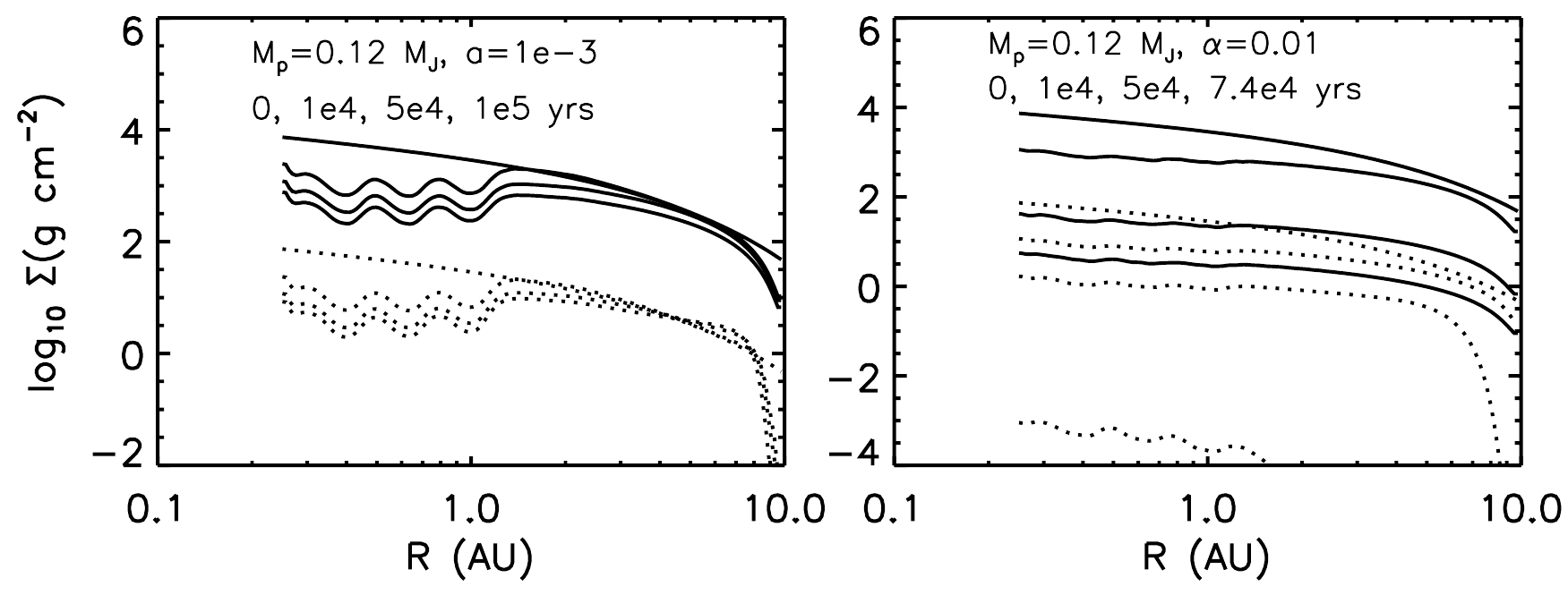}
 \caption{Hydrodynamical simulations of the surface density (solid 
line=gas; dotted line=1 mm dust) at different times for viscous disks 
($\alpha$=0.001, left; $\alpha$=0.01, right) with three accreting 
$0.12\,M_J$ planets, at radii 0.4, 0.63, and 1 au.}
 \label{Fig2}
 \end{figure}

\clearpage

\begin{figure}
\includegraphics[height=7.2cm]{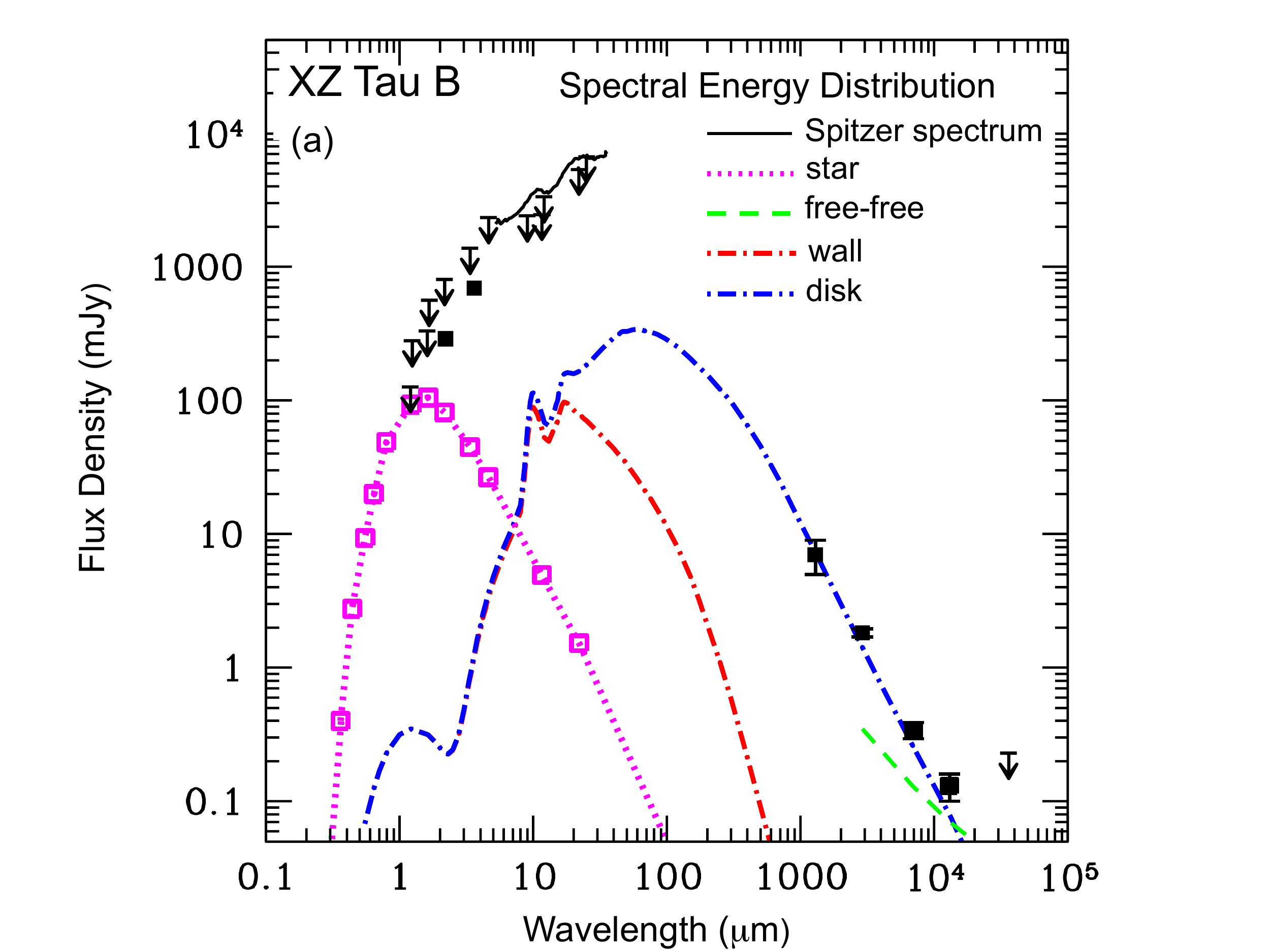}
\includegraphics[height=6.2cm]{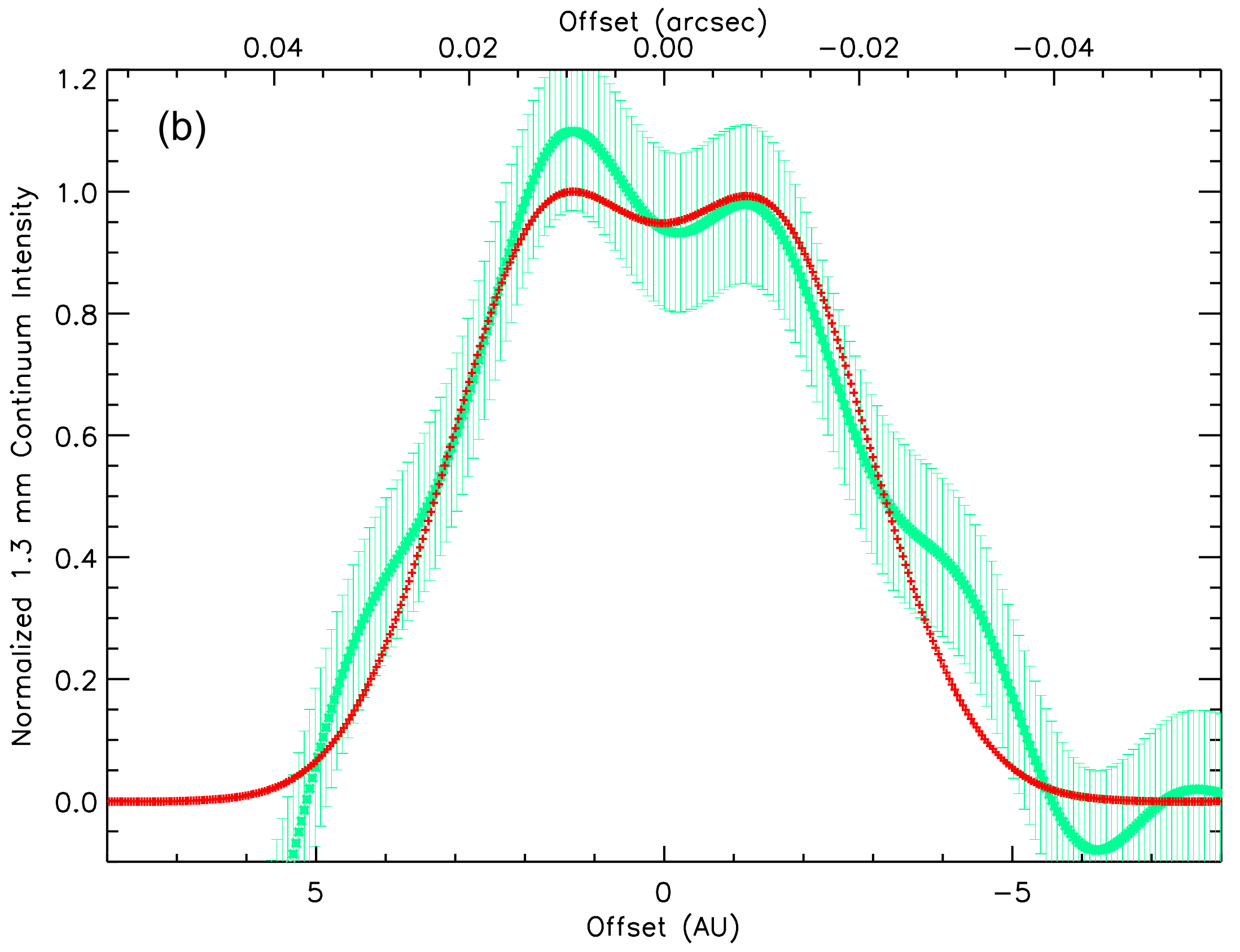}
\includegraphics[height=7.5cm]{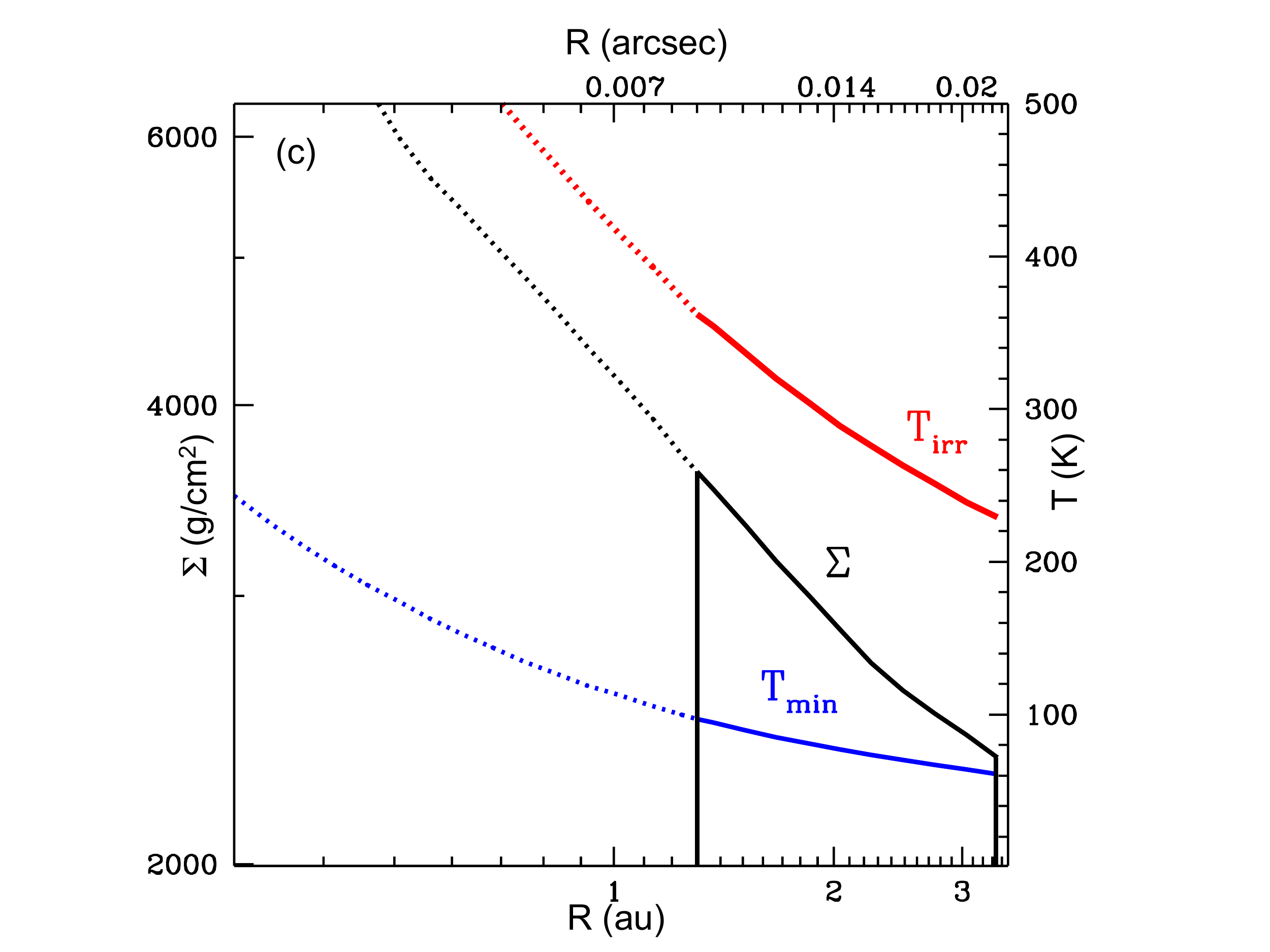}
 \hfil
\includegraphics[height=8.2cm]{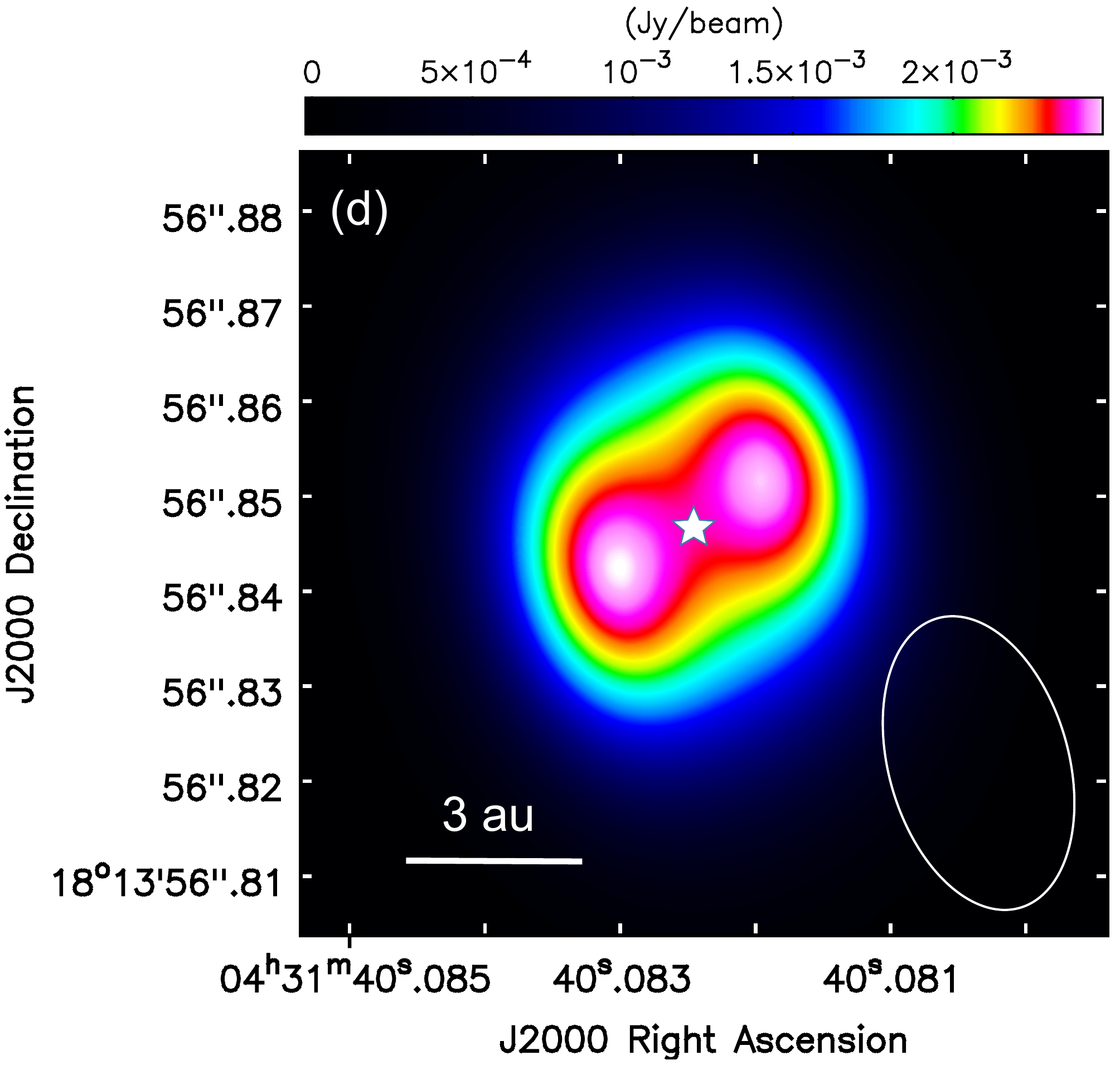}
 \caption{\small (a) Spectral energy distribution. Solid square symbols 
and continuous line are observational data. Arrows are upper limits, 
corresponding to data that do not separate XZ Tau B from A/C components. 
Open squares represent the calculated contribution of the star (see 
text). Discontinuous lines represent the model results. (b) Observed 
(green line with 1-$\sigma$ error bars) and model (red line) normalized 
intensity profile along a line passing through the two maxima 
(PA=126$^\circ$) of the disk (positive offsets to the east). (c) Model 
surface density ($\Sigma$), irradiation temperature ($T_{\rm irr}$), and 
minimum temperature ($T_{\rm min}$) radial distributions. (d) CASA 
simulated image of the 1.3 mm model emission, assuming the same antenna 
configuration as in Figure 1a.}
 \label{Fig3}
 \end{figure}

\clearpage

\begin{deluxetable}{lccc}
\tablewidth{0pt}
\tablecaption{Parameters of the XZ Tau B Star and Disk\label{Tab1}}
\tablehead{
\colhead{Parameter} &
\colhead{Value} &
\colhead{Notes} &
\colhead{Refs.}
}
\startdata
\multicolumn{4}{c}{Star} \\
\hline
Distance (pc)  & 140 & Adopted & 1\\
Visual Extinction (mag) & 1.4 & Adopted  & 2\\
Spectral Type & M2 & Adopted  & 2\\
Age (Myr) & 4.6 & Adopted  & 2\\
Effective Temperature (K) & 3550 &  Adopted& 2\\
Radius ($R_{\odot})$  & 1.24 & Calculated & \\
Mass  ($M_{\odot}$)   & 0.37 & Adopted  & 2\\
Mass Accretion Rate ($M_{\odot}$ yr$^{-1}$) & 1.4$\times$10$^{-8}$ & Calculated &\\
\hline 
\multicolumn{4}{c}{Disk} \\
\hline 
Inclination Angle (deg) & 35$\pm$10 & Adopted/Refined  & 3\\
Position Angle of Major Axis (deg) & 140$\pm$10 & Adopted/Refined  & $4$ \\
Inner Radius (au)  & 1.30$\pm$0.05  & Fitted\\
Outer Radius (au)  & 3.4$\pm$0.1  & Fitted\\ 
Viscosity Parameter & 0.001 & Adopted\\
Mass Accretion Rate ($M_{\odot}$ yr$^{-1}$) & 7.0$\times$10$^{-8}$ & Fitted& \\
Degree of Settling & 0.10  & Fitted \\
1.3 mm Optical Depth at 1 au &18 & Calculated \\
1.3 mm Optical Depth at 3.4 au & 14 & Calculated \\
Mass Evacuated in Cavity ($M_J$) & 3 & Calculated  \\
Total Mass ($M_J$)  & 9 & Calculated\\
Dust Mass ($M_\oplus$)  & 25 & Calculated\\
\hline 
\multicolumn{4}{c}{Cavity Wall} \\
\hline 
Radius (au)  & 1.30$\pm$0.05 &= Disk Inner Radius\\
Temperature (K) & 420 & Calculated\\
Height (au) & 0.09 & = Disk Hydrostatic Scale Height\\
\enddata
\tablerefs{(1) Torres et al. 2009; (2) Hartigan \& Kenyon 2003; (3) Carrasco-Gonz\'alez et al. in preparation (4) Krist et al. 2008}
\end{deluxetable}

\clearpage

\acknowledgments


Support from MINECO-FEDER AYA2014-57369-C3 grant, CONACyT, and 
DGAPA-UNAM is acknowledged. This paper makes use of the following ALMA 
data: ADS/JAO.ALMA\#2011.0.00015.SV. ALMA is a partnership of ESO 
(representing its member states), NSF (USA) and NINS (Japan), together 
with NRC (Canada), NSC and ASIAA (Taiwan), and KASI (Republic of Korea), 
in cooperation with the Republic of Chile. The Joint ALMA Observatory is 
operated by ESO, AUI/NRAO and NAOJ. The National Radio Astronomy 
Observatory is a facility of the National Science Foundation operated 
under cooperative agreement by Associated Universities, Inc.

{\it Facilities:} \facility{ALMA}


\begin{thebibliography}{}



\bibitem[ALMA Partnership et al.(2015)]{2015ApJ...808L...3A} ALMA 
Partnership, Brogan, C.~L., P{\'e}rez, L.~M., et al.\ 2015, \apjl, 808, 
L3 (AP2015a)

\bibitem[ALMA Partnership et al.(2015)]{2015ApJ...808L...1A} ALMA 
Partnership, Fomalont, E.~B., Vlahakis, C., et al.\ 2015, \apjl, 808, L1 
(AP2015b)

\bibitem[Andrews(2015)]{2015PASP..127..961A} Andrews, S.~M.\ 2015, 
\pasp, 127, 961 

\bibitem[Andrews et al.(2011)]{2011ApJ...732...42A} Andrews, S.~M., Wilner, 
D.~J., Espaillat, C., et al.\ 2011, \apj, 732, 42 

\bibitem[Andrews et al.(2009)]{2009ApJ...700.1502A} Andrews, S.~M., 
Wilner, D.~J., Hughes, A.~M., Qi, C., \& Dullemond, C.~P.\ 2009, \apj, 
700, 1502

\bibitem[Birnstiel et al.(2013)]{2013A&A...550L...8B} Birnstiel, T., 
Dullemond, C.~P., \& Pinilla, P.\ 2013, \aap, 550, L8

\bibitem[Blum \& Wurm(2008)]{2008ARA&A..46...21B} Blum, J., \& Wurm, G.\ 
2008, \araa, 46, 21

\bibitem[Calvet et al.(2005)]{2005ApJ...630L.185C} Calvet, N., 
D'Alessio, P., Watson, D.~M., et al.\ 2005, \apjl, 630, L185

\bibitem[Calvet \& Gullbring(1998)]{1998ApJ...509..802C} Calvet, N., \& 
Gullbring, E.\ 1998, \apj, 509, 802 

\bibitem[Carrasco-Gonz{\'a}lez et al.(2009)]{2009ApJ...693L..86C} 
Carrasco-Gonz{\'a}lez, C., Rodr{\'{\i}}guez, L.~F., Anglada, G., 
\& Curiel, S.\ 2009, \apjl, 693, L86 

\bibitem[Coffey et al.(2004)]{2004A&A...419..593C} Coffey, D., Downes, 
T.~P., \& Ray, T.~P.\ 2004, \aap, 419, 593 

\bibitem[D'Alessio et al.(2006)]{2006ApJ...638..314D} D'Alessio, P., 
Calvet, N., Hartmann, L., Franco-Hern{\'a}ndez, R., 
\& Serv{\'{\i}}n, H.\ 2006, \apj, 638, 314 

\bibitem[D'Alessio et al.(2005)]{2005ApJ...621..461D} D'Alessio, P., 
Hartmann, L., Calvet, N., et al.\ 2005, \apj, 621, 461 

\bibitem[de Gregorio-Monsalvo et al.(2013)]{2013A&A...557A.133D} de 
Gregorio-Monsalvo, I., M{\'e}nard, F., Dent, W., et al.\ 2013, \aap, 
557, A133

\bibitem[Dupuy et al.(2016)]{2016ApJ...817...80D} Dupuy, T.~J., Kratter, 
K.~M., Kraus, A.~L., et al.\ 2016, \apj, 817, 80 

\bibitem[Dzib(2014)]{2014ApJ...788..162D} Dzib, S.~A., Loinard, L., 
Rodr{\'i}guez, L. F., Galli, P.\ 2014, \apj, 788, 162

\bibitem[Espaillat et al.(2008)]{2008ApJ...682L.125E} Espaillat, C., 
Calvet, N., Luhman, K.~L., Muzerolle, J., \& D'Alessio, P.\ 2008, \apjl, 
682, L125

\bibitem[Espaillat et al.(2014)]{2014prpl.conf..497E} Espaillat, C., 
Muzerolle, J., Najita, J., et al.\ 2014, Protostars and Planets VI, 497

\bibitem[Forgan et al.(2014)]{2014MNRAS.439.4057F} Forgan, D., Ivison, 
R.~J., Sibthorpe, B., Greaves, J.~S., \& Ibar, E.\ 2014, \mnras, 439, 4057 

\bibitem[Furlan et al.(2016)]{2016ApJS..224....5F} Furlan, E., Fischer, 
W.~J., Ali, B., et al.\ 2016, \apjs, 224, 5

\bibitem[Hartigan \& Kenyon(2003)]{2003ApJ...583..334H} Hartigan, P., \& 
Kenyon, S.~J.\ 2003, \apj, 583, 334

\bibitem[Hioki et al.(2009)]{2009PASJ...61.1271H} Hioki, T., Itoh, Y., 
Oasa, Y., et al.\ 2009, \pasj, 61, 1271 

\bibitem[Hughes et al.(2007)]{2007ApJ...664..536H} Hughes, A.~M., 
Wilner, D.~J., Calvet, N., et al.\ 2007, \apj, 664, 536

\bibitem[Jontof-Hutter et al.(2015)]{2015Natur.522..321J} Jontof-Hutter, 
D., Rowe, J.~F., Lissauer, J.~J., Fabrycky, D.~C., \& Ford, E.~B.\ 2015, 
\nat, 522, 321

\bibitem[Kenyon \& Hartmann(1995)]{1995ApJS..101..117K} Kenyon, S.~J., 
\& Hartmann, L.\ 1995, \apjs, 101, 117 

\bibitem[Kraus et al.(2015)]{2015ApJ...798L..23K} Kraus, A.~L., Andrews, 
S.~M., Bowler, B.~P., et al.\ 2015, \apjl, 798, L23 

\bibitem[Krist et al.(2008)]{2008AJ....136.1980K} Krist, J.~E., 
Stapelfeldt, K.~R., Hester, J.~J., et al.\ 2008, \aj, 136, 1980 

\bibitem[Lee et al.(2014)]{2014ApJ...797...95L} Lee, E.~J., Chiang, E., 
\& Ormel, C.~W.\ 2014, \apj, 797, 95 

\bibitem[Lissauer et al.(2014)]{2014Natur.513..336L} Lissauer, J.~J., 
Dawson, R.~I., \& Tremaine, S.\ 2014, \nat, 513, 336 

\bibitem[Lissauer et al.(2011)]{2011Natur.470...53L} Lissauer, J.~J., 
Fabrycky, D.~C., Ford, E.~B., et al.\ 2011, \nat, 470, 53 

\bibitem[McClure et al.(2015)]{2015ApJ...799..162M} McClure, M.~K., 
Espaillat, C., Calvet, N., et al.\ 2015, \apj, 799, 162 

\bibitem[McClure et al.(2008)]{2008ApJ...683L.187M} McClure, M.~K., 
Forrest, W.~J., Sargent, B.~A., et al.\ 2008, \apjl, 683, L187 

\bibitem[Ogihara et al.(2015)]{2015A&A...578A..36O} Ogihara, M., 
Morbidelli, A., \& Guillot, T.\ 2015, \aap, 578, A36

\bibitem[Osorio et al.(2014)]{2014ApJ...791L..36O} Osorio, M., Anglada, G., 
Carrasco-Gonz{\'a}lez, C., et al.\ 2014, \apjl, 791, L36 

\bibitem[Papaloizou et al.(2007)]{2007prpl.conf..655P} Papaloizou, 
J.~C.~B., Nelson, R.~P., Kley, W., Masset, F.~S., 
\& Artymowicz, P.\ 2007, Protostars and Planets V, 655 

\bibitem[Papaloizou \& Pringle(1977)]{1977MNRAS.181..441P} Papaloizou, 
J., \& Pringle, J.~E.\ 1977, \mnras, 181, 441

\bibitem[Pecaut \& Mamajek(2013)]{2013ApJS..208....9P} Pecaut, M.~J., \& 
Mamajek, E.~E.\ 2013, \apjs, 208, 9 

\bibitem[Pi{\'e}tu et al.(2014)]{2014A&A...564A..95P} Pi{\'e}tu, V., 
Guilloteau, S., Di Folco, E., Dutrey, A., \& Boehler, Y.\ 2014, \aap, 
564, A95

\bibitem[Rodr{\'{\i}}guez et al.(1998)]{1998Natur.395..355R} 
Rodr{\'{\i}}guez, L.~F., D'Alessio, P., Wilner, D.~J., et al.\ 1998, \nat, 
395, 355 

\bibitem[Ros \& Johansen (2013)]{2013A&A.552..A137} Ros, K., Johansen, 
A.\ 2013, \aap, 552, A137

\bibitem[Sandell \& Aspin(1998)]{1998A&A...333.1016S} Sandell, G., \& 
Aspin, C.\ 1998, \aap, 333, 1016

\bibitem[Siess et al.(2000)]{2000A&A...358..593S} Siess, L., Dufour, E., 
\& Forestini, M.\ 2000, \aap, 358, 593 

\bibitem[Swift et al.(2013)]{2013ApJ...764..105S} Swift, J.~J., Johnson, 
J.~A., Morton, T.~D., et al.\ 2013, \apj, 764, 105 

\bibitem[Torres et al.(2009)]{2009ApJ...698..242T} Torres, R.~M., Loinard, 
L., Mioduszewski, A.~J., \& Rodr{\'{\i}}guez, L.~F.\ 2009, \apj, 698, 242 

\bibitem[van der Marel et al.(2013)]{2013Sci...340.1199V} van der Marel, 
N., van Dishoeck, E.~F., Bruderer, S., et al.\ 2013, Science, 340, 1199 

\bibitem[White \& Ghez(2001)]{2001ApJ...556..265W} White, R.~J., \& 
Ghez, A.~M.\ 2001, \apj, 556, 265

\bibitem[Williams \& Cieza(2011)]{2011ARA&A..49...67W} Williams, J.~P., 
\& Cieza, L.~A.\ 2011, \araa, 49, 67 

\bibitem[Zapata et al.(2015)]{2015ApJ...811L...4Z} Zapata, L.~A., 
Galv{\'a}n-Madrid, R., Carrasco-Gonz{\'a}lez, C., et al.\ 2015, \apjl, 
811, L4 

\bibitem[Zhu et al.(2012)]{2012ApJ...755....6Z} Zhu, Z., Nelson, R.~P., 
Dong, R., Espaillat, C., \& Hartmann, L.\ 2012, \apj, 755, 6 

\bibitem[Zhu et al.(2011)]{2011ApJ...729...47Z} Zhu, Z., Nelson, R.~P., 
Hartmann, L., Espaillat, C., \& Calvet, N.\ 2011, \apj, 729, 47 

\end{thebibliography}
\end{document}